\documentclass[galaxies,article,accept,oneauthor,pdftex,10pt,a4paper]{mdpi}

\newcommand{\ssr}{Space~Sci.~Rev.}	
\newcommand{\nat}{Nature}		
\firstpage{1}
\articlenumber{x}
\doinum{10.3390/------}
\pubvolume{xx}
\pubyear{2018}
\copyrightyear{2018}
\history{Received: 29 January 2018; Accepted: 20 March 2018; Published: date}

\usepackage{booktabs}
\usepackage{multirow}
\usepackage{soul}
\usepackage{microtype}

\usepackage{upgreek}
\usepackage{textcomp}

\Title{Future X-ray Polarimetry of Relativistic Accelerators: Pulsar Wind Nebulae and Supernova Remnants}

\Author{Niccol\`o Bucciantini $^{1,2,3}$\orcidA{}}

\AuthorNames{Firstname Lastname, Firstname Lastname and Firstname Lastname}

\address{%
$^{1}$ \quad INAF---Osservatorio Astrofisico di Arcetri, Largo E. Fermi 5,
I-50125 Firenze, Italy; niccolo@arcetri.astro.it\\
$^{2}$ \quad Dipartimento di Fisica e Astronomia, Universit\`a degli Studi di Firenze, Via G. Sansone 1,
 Sesto F.~no, I-50019~Firenze, Italy\\
$^{3}$ \quad INFN---Sezione di Firenze, Via G. Sansone 1, Sesto F.~no, I-50019 Firenze, Italy}

\corres{Correspondence: niccolo@arcetri.astro.it; Tel.: +39-055-2752-285
\textcolor{red}{(Please carefully check the accuracy of names and affiliations. Changes will not be possible after proofreading.)}}

\abstract{Supernova remnants (SNRs) and pulsar wind nebulae (PWNs) are among the most significant sources of
  non-thermal X-rays in the sky, and the best means by which 
  relativistic plasma dynamics and particle acceleration can be
  investigated. Being strong synchrotron emitters, they are ideal
  candidates for X-ray polarimetry, and indeed the Crab nebula is up
  to present the only object where X-ray polarization has been
  detected with a high level of  significance.  Future polarimetric
  measures will likely provide us with crucial information on the level of
  turbulence that is expected at   particle acceleration sites,
  together with the spatial and temporal coherence of   magnetic
  field geometry, enabling us to set stronger constraints on our
  acceleration models.  PWNs  
   will also allow us to estimate the
  level of internal dissipation. I will briefly review the current  knowledge on the polarization signatures in SNRs and PWNs, and I will illustrate what  we can hope to achieve with future missions such as IXPE/XIPE. 
  }


\begin{document}


\section{Introduction}

Pulsar wind nebulae (PWNs)
are bubbles of relativistic particles (mostly pairs) and
magnetic fields that form when  the relativistic pulsar
wind interacts with the ambient medium (interstellar medium (ISM) 
 or supernova remnant (SNR)). They shine in
non-thermal (synchrotron and inverse Compton) radiation
 in a broad range of frequencies from radio wavelengths to
$\gamma$-rays (see \cite{Gaensler_Slane06a} for a review).
PWNs are at present one of the more promising astrophysical environments
where relativistic outflows and relativistic shock acceleration can
be investigated. They are, above all, one of the most efficient antimatter
factories present in the galaxy and have been advocated as a possible
source of the so-called ``positron excess'' \cite{Adriani_Barbarino+09a,Blasi_Amato11a}.

At X-rays, many PWNs exhibit an axisymmetric feature known as a
 {\it jet-torus structure}. This feature has been observed by now in a
number of PWNs, among which are the Crab nebula \cite{Weisskopf_Hester+00a},
Vela \cite{Pavlov_Kargaltsev+01a}, and MSH~15-52
\cite{Gaensler_Arons+02a}, to name just a few.
It is now commonly accepted that this
structure arises due to the interplay between the anisotropic energy flux
in the wind and the compressed toroidal magnetic field in the nebula, 
as confirmed by a long
series of numerical simulations
\cite{Komissarov_Lyubarsky04a,Del-Zanna_Amato+04a,Olmi_Del-Zanna+16a},
and that its shape and properties can be used to probe the structure
of the otherwise unobservable pulsar wind, and the acceleration
properties of the wind termination shock \cite{Olmi_Del-Zanna+15a}.

Shell SNRs trace the ejected layers of the parent
star, launched during the supernova explosion, as they propagate
into the ISM, driving a  high Mach number forward shock where the ambient matter is heated
and compressed, and particles are accelerated \cite{Woltjer72a,Reynolds08a}. While~the stellar
ejecta are mostly revealed as thermal line emission, the forward
shock is seen as a bright non-thermal limb, shining in synchrotron from
radio waves to X-rays. SNRs are commonly thought to be   at the origin of
the bulk of the galactic cosmic rays (CRs): diffusive shock
acceleration, consisting of repeated crossings of the surface of the
shock, is capable of rising the particle energy up to \mbox{\textasciitilde{}0.1--1
PeV \cite{Reynolds11a}}.

In the last decade, it has become clear that the acceleration process
can strongly modify the dynamics of the shock, and in particular can
  substantially amplify the magnetic field upstream of the shock
itself, driving the development of magnetic turbulence \cite{Bell04a,Reynolds_Gaensler+12a,Mattheus_Bell+17a,Xu_Lazarian17a}. This has
crucial implications for the particle spectra and the maximum energy
that can be achieved. The synchrotron X-rays seen in young  SNRs are
due to accelerated electrons with typical energies of 1--10 TeV in a
magnetic field of a few hundred  $\upmu$G \cite{Reynolds_Gaensler+12a}. Such high values of
the magnetic field strength cannot be explained by shock compression alone,
but can be produced by instabilities in the upstream due to the accelerated particles
themselves \cite{Bell04a}.

\section{Radio \& Optical Polarization}

Radio polarimetry of PWNs and SNRs has a long history. In PWNs, as radio emission
is dominated by the outer regions of the nebula, where the effects of the
interaction with the SNR are stronger, and where Rayleight--Taylor
instability operates, radio polarimetry provides at best an estimate of the
degree of ordered versus disordered magnetic field. This means that it cannot be used to probe the conditions in the region close to the
termination shock, where most of the variability and the acceleration
processes take place. In the Crab nebula, which constitutes a case
study for the entire class, the radio polarized fraction is
\textasciitilde{}16\% on average \cite{Conway71a,Ferguson73a,Velusamy85a,Aumont_Conversi+10a} with
peaks up to $30\%$, which is lower than the average optical polarized 
fraction, which  is \textasciitilde{}25\% \cite{Velusamy85a}. 
Interestingly, the polarized
flux in radio anti-correlates with the position of the bright X-ray
torus.

In other systems, the interpretation of the radio polarized morphology
can be quite challenging. Vela shows a clear toroidal pattern,
consistent with the orientation of the double ring that is observed in
X-rays \cite{Dodson_Lewis+03a}. A similar highly ordered toroidal pattern is seen
also in G106.6-29 \cite{Kothes_Reich+06a}. This is consistent with the general
expectation of a synchrotron bubble where a highly wound-up  magnetic field is
inflated by the wind coming from a rapid rotator. Other systems clearly
show a far more complex morphology, ranging from a highly turbulent
structure \cite{Ma_Ng+16a}, typical in old systems that have gone through a strong
interaction phase with the SNR known as the reverberation phase
\cite{Blondin_Chevalier+01a,Bucciantini_Blondin03a}, to one that is mostly radial (or dipole-like) 
  \cite{Kothes_Landecker+08a}. There is at
the moment no framework to interpret these differences or to relate
them consistently to the dynamics of the PWNs.

For the Crab nebula, high resolution HST 
observations in polarized
light for the inner region, in~particular  the brightest
optical features, namely the {\it
  knot} and the {\it wisps}, show typical polarized
fractions of about 60\% and 40\%, respectively \cite{Moran_Shearer+13b}. The results in
the  Crab nebula are
consistent  with the general idea of a mostly toroidal
magnetic field just downstream of the termination shock, with~a
possible hint of developing turbulence: the polarized fraction
of the wisps is lower than the one in the knot, and at present,
emission maps based on numerical simulations suggest that the former is slightly more downstream than
the latter. It is interesting to notice that, while it is the brightest
feature in total light, the torus has a lower surface brightness than
the wisp in polarized light \cite{Hester08a}.  There is no optical
counterpart to Vela, neither in total nor in polarized light \cite{Marubini_Sefako+15a,Moran_Mignani+14a}.
As of today, optical polarization is limited to the brightest features of the
brightest nebula. Moreover, the optical light 
 usually suffers from large foreground
contamination and is often polarized, 
 and the jet-torus structure, which is
substantially prominent in X-rays, is much fainter.

The radio polarization of shell SNRs shows an interesting dichotomy
between young systems, where the magnetic field appears to be
predominantly radial, and old systems, where it looks tangential to the
shock front \cite{Dubner_Giancani15a}. This is commonly interpreted as an evidence
for a stronger level of instability in young and more energetic
systems, more likely related to the formation of Rayleigh--Taylor
\mbox{fingers  \cite{Jun98a,Bucciantini_Amato+04a}} at the contact discontinuity between the shocked ISM and the
shocked SN ejecta, which will act to preferentially stretch the field
in the radial direction. Alternative models invoking other kinds of
instabilities, such as Richtmyer--Meshkov, have also been presented
\cite{Inoue_Shimoda+13a}, as well as models with magnetic dependent acceleration \cite{West_Jaffe+17a}. In older systems, where these instabilities are
supposed to be less effective, the field has the naive geometry that one
would expect from shock compression (note that shock compression will
amplify the tangential component of the field, producing a strong
polarization pattern even if the upstream field is strongly
turbulent). This dichotomy appears independently of the progenitor type
of the SNR.

On the other hand, there are lines of evidence suggesting that some
signatures of the upstream mean field are preserved. In particular, there
appears to be a correlation between the orientation of bipolar SNRs and the
galactic plane 
\cite{Gaensler98a,Reynoso_Hughes+13a}, and there is further evidence in  SNRs from Type II SN
of an expansion into a magnetized wind bubble \cite{Harvey-Smith_Gaensler+10a}. It was found in SN 1006
that the polarized fraction anti-correlates with radio emissions,
suggesting that those sites along the shock front that are more likely
to accelerate particles  have   a more turbulent field \cite{Reynoso_Hughes+13a}. It is well
known that the level of turbulence and the orientation of the field
are pivotal for particle acceleration models: perpendicular shocks are
thought to be more efficient accelerators, while parallel shocks tend
to be more efficient injectors. Moreover, turbulence is likely
required to explain the high magnetic field strength required in SED 
fitting of shell SNRs \cite{Reynolds08a}. As for PWNs, radio
polarization in SNRs traces particles with lifetimes that are longer than the age
of the nebulae that are filling the shell volume. 
In Cas A, for
example, high resolution radio observations show evidence of a
polarization angle swing at the location of the X-ray rim, which is where~particles are accelerated \cite{Gotthelf_Koralesky+01a}.

\section{Polarization Models}

In the last decade, several multidimensional models have been put forward
to investigate the magnetic field structure and the geometry of the
flow in PWNs. Much of the work has focused on trying to reproduce the
jet-torus structure and to use it as a probe for the properties of the
otherwise unobservable pulsar wind \cite{Komissarov_Lyubarsky04a,Del-Zanna_Amato+04a,Olmi_Del-Zanna+16a}. These works have shown
the importance of Doppler boosting effects and have enabled us to
locate the possible origin on many of the primary axisymmetric
features observed in PWNs, including the {\it knot} and {\it wisps} of the
Crab nebula. On the other hand, in present day numerical models, the
torus tends to be under-luminous with respect to the wisps, and despite
being a strong dynamical feature, the jet is hard to reproduce. What
is missing in current day models is the possible presence of magnetic
turbulence, at scales that are too small to be resolved by our numerical tools but are sufficiently large to affect emission. Several theoretical
arguments  suggesting that a non-negligible amount of magnetic turbulence should be present in PWNs have been put forward in recent years. For example, the presence of    diffuse X-ray halos has been stated to be
larger than what is expected  for synchrotron
cooling and advection \cite{Tang_Chevalier12a,Buhler_Blandford14a,Zrake_Arons16a}, it has been suggested that radio-emitting
particles are accelerated in the bulk of the
nebula  \cite{Olmi_Del-Zanna+15a,Tanaka_Asano16a}, and the recurrent $\gamma$-ray flares have been
 interpreted as dissipation in localized strong current sheets \cite{Uzdensky_Cerutti+11a}. 
 The possible origin of
such turbulence is unclear: it could simply be the magnetic cascade of
the large-scale turbulence injected at the
termination shock \cite{Camus_Komissarov+09a}; it could be due to residual reconnection
taking place downstream of the shock in a striped wind \cite{Sironi_Spitkovsky11a}; it also
could be related to current-driven instability of the compressed
toroidal field \cite{Mizuno_Lyubarswky+11a,O'neill_Beckwith+12a}.

Based on the idea that small-scale turbulence can be present, we have
developed a formalism to include it, as a sub-grid effect \cite{Bandiera_Petruk16a,Petruk_Bandiera+17a,Bucciantini_Bandiera+17a}, into large
scale models for the global structure of the field, either simplified
toy models on the line of \cite{Ng_Romani04a}, which are easy and fast to
compute and allow us to deeply scan the possible parameter space in
order to optimize the agreement with observations, or~more sophisticated
time-dependent numerical models  that can take into account the
interplay between the pulsar wind and the environment. These models have
been recently applied to the Crab and Vela nebulae~\cite{Bucciantini_Bandiera+17a}.  It has been shown
that, in order to recover the current-relative 
brightness between the wisps
and torus as well as the correct luminosity profile of the torus in
Crab and of the inner and outer ring in Vela, a~substantial fraction
of magnetic energy of \textasciitilde{}50\% must be in the form of a turbulent small-scale field. This number  represents   a typical integrated polarized
fraction for the Crab nebula of \textasciitilde{}17\%, consistent with existing
measures.

SNRs have a long history of polarization models for the radio band, and such models
  have attempted to constrain the origin of the observed
polarization dichotomy, including recipes to relate it to the
physics of acceleration \cite{Reynolds_Gaensler+12a,West_Jaffe+17a}. More recently, the same technique used to
include the effects of a turbulent component in the emission of PWNs
have been applied to shell SNRs, trying to derive possible
observational constraints to locate the regions where the turbulence is
higher, and to assess its correlation with particle acceleration
sites \cite{Petruk_Bandiera+17a}.  One of the most interesting aspects of X-ray emission
in SNRs  is that it takes place close to the cut-off regime: this
implies that emission tends to weigh     regions of higher magnetic
fields to a greater extent, and this means that large differences in the polarized
emission pattern are expected for shallow vs. steep  magnetic
turbulent spectra. The authors of \cite{Bykov_Uvarov+09a} have used a simplified model that takes
into account the typical emissivity expected in shell SNRs, to
evaluate the level and structure of polarized emission expected from
different turbulent spectra, and found that even the simple detection
of a polarized signal is enough to rule out the shallower cascades.    

More interestingly, a polarized emission model has been recently
presented to explain the striped zone observed in X-rays  in Tycho
SNR \cite{Bykov_Ellison+11a}. It has been suggested that such stripes might trace turbulent
magnetic fields generated by accelerated particles streaming upstream
of the the shock itself. The~orientation of the magnetic field with
respect to the stripes might enable us to constrain the kind of
instability driving the amplification of the field, given that
different mechanisms provide different polarization patterns \cite{Bell04a,Bykov_Osipov+11a}.

\section{Prospects for Future Observations}

Ideally one would like to probe these systems using X-ray
polarimetry \cite{Bucciantini10a}, and there is great interest among the scientific
community for such an objective \cite{Soffitta_Barcons+13a,Weisskopf_Ramsey+16a}. Incidentally, the Crab nebula
is at the moment the only object with a polarization detected in X-rays
\cite{Weisskopf_Silver+78a}. The Crab nebula has been more recently
observed in X-ray polarized light by NuSTAR \cite{Madsen_Reynolds+15a}
and by  PoGO+
\cite{Chauvin_Floren+16a,Chauvin_Floren+17a}. High-energy measures by INTEGRAL are also available
\cite{Forot_Laurent+08a}, and there are suggestions of a possible  time variation in the polarization
angle \cite{Moran_Kyne+16a}.  Recent polarized detection at high
energies has also been reported by AstroSat \cite{Vadawale_Chattopadhyay+18a}.

In recent years, a renewed interest in modeling space-resolved polarimetric
measures has come as a result of the great efforts made to develop the
IXPE and XIPE missions 
\cite{Soffitta_Barcons+13a,Weisskopf_Ramsey+16a}. Simulations using the baseline combined telescope effective area and
point spread function (PSF) were performed for both~instruments.

\begin{figure}[H]
\centering
\includegraphics[width=12 cm]{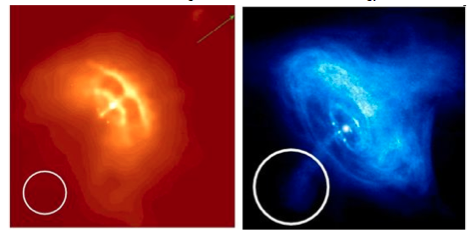}
\caption{\textls[-15]{Chandra images of the Vela (\textbf{left}) and Crab (\textbf{right}) PWNs. The
  IXPE 30'' resolution (half power diameter) is shown in the lower
  left corners of the images. From the IXPE Science-Investigation~Document.}}
  \label{fig1}
\end{figure}

For Crab and Vela, IXPE will be able to measure the polarization in a
number of different spatial resolution elements, thus providing the
first spatially resolved X-ray polarimetry of a PWN (Figure~\ref{fig1}). For Crab, it is
estimated that a 7.3-day 
 observation can detect a polarized fraction well
below 2\% at 99\% confidence in each of five distinct spatial regions,
including one centered on the jet. This takes into account the fact
that 50\% of the flux might originate in neighboring zones and be
unpolarized.
For~Vela, a polarization of the entire nebula of 3\% may be detected in
a 4.6-day observation, allowing also for some spatially resolved imaging with
a higher threshold of \textasciitilde{}5--10\%.
For other bright  PWNs powered by   young pulsars such as PSR B1509-58
and J1833-1034, it will be easy with a few days of observation to have
enough statistics to obtain an integrated polarized fraction and, perhaps
in the case of B1509-58,   to also obtain the polarization of the bright jet.

\begin{figure}[H]
\centering
\includegraphics[width=15 cm]{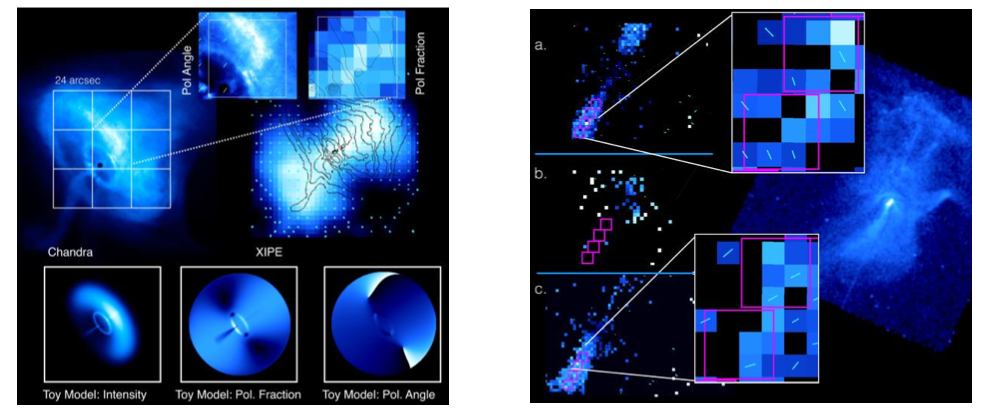}
\caption{Left panel ({\bf {a}}): Simulation of the Crab Nebula as seen
  by XIPE in 0.2 Msec. The toy model mimics the Chandra image for a
  given polarization angle and fraction. Right panel ({\bf b}):
  Simulations of the PWN MSH 15-52 as seen by XIPE in 2 Msec. The
  images a, b, and c show the expected intensity map when
  three different polarization models are applied to the Chandra intensity map:
   (\textbf{a}) a fully ordered radial B-field, (\textbf{b}) a fully disordered
  B-field, and (\textbf{c}) a fully ordered perpendicular B-field. From the
  XIPE Yellow Book, see also \cite{Wilhelmi_Vink+17a}.}
    \label{fig2}
\end{figure}

   XIPE  various simulations of  different
scenarios with   magnetic field orientations based on simplified toy
models \cite{Bucciantini_Bandiera+17a} (see Figure~\ref{fig2}) were
carried out both for Crab and MSH 15-52 \cite{Wilhelmi_Vink+17a} and, with just a 0.2 ksec observation for Crab and a 2 Msec
observation for MSH~15-52, 
 showed that the
polarized patterns are reconstructed with errors
of less than 0.1\% and with 
more than 10$\sigma$ within the instrument
PSF   (see~the contribution by J. Vink in these
same proceedings for   details on the modeling of these observations, the
instrumental response, and the robustness of the results).

The main targets for polarization measures among shell SNRs  are the few that are
  young and close-by,   have enough surface brightness, and are large
enough to be resolved. 
XIPE can detect polarized X-ray emission from
Tycho's SNR with enough resolution to allow us to set
constraints on models of diffusive shock acceleration with efficient
magnetic field amplification in SNRs, with a typical integration time of
\textasciitilde{}1 Msec. In Tycho, the emission in the 4--6 keV range is expected to
be of synchrotron origin. The current Monte Carlo method of constructing
simulated observation are based on   theoretical models for polarized
emission constructed by \cite{Wilhelmi_Vink+17a}.

Another primary target among SNRs will be Cas A, where thermal line
emission associated with the interior filaments is present. A good
energy resolution is pivotal to select those parts of the emitted
X-ray radiation that are of non-thermal origin (far from the
lines). In Cas A, X-ray emission is detected also from the putative
reverse shock. Simulated observations suggest that, with  XIPE, it is
possible to disentangle the polarization signature of the reverse shock
from that of the forward shock, as long as typical values are \textasciitilde{}10\% (see the contribution by J. Vink in these same proceedings).
Among   other possible targets, there are  SN 1006, RX J1713.7-394,  Kepler's
SNR, and RCW 86.

\unskip

\conflictsofinterest{The authors declare no conflicts of interest.\textcolor{red}{(Please confirm the conflict of interest.)}}


\reftitle{References}





\begin{thebibliography}{-------}
\providecommand{\natexlab}[1]{#1}

\end{thebibliography}


\begin{thebibliography}{999}

\bibitem[{Gaensler} and {Slane}(2006)]{Gaensler_Slane06a}
{Gaensler}, B.M.; {Slane}, P.O.
\newblock {The Evolution and Structure of Pulsar Wind Nebulae}.
\newblock {\em Annu. Rev. Astron. Astrophys.} {\bf 2006}, {\em 44},~17--47.

\bibitem[{Adriani} \em{et~al.}(2009){Adriani}, {Barbarino}, {Bazilevskaya},
  {Bellotti}, {Boezio}, {Bogomolov}, {Bonechi}, {Bongi}, {Bonvicini}, {Bottai},
  {Bruno}, {Cafagna}, {Campana}, {Carlson}, {Casolino}, {Castellini}, {de
  Pascale}, {de Rosa}, {de Simone}, {di Felice}, {Galper}, {Grishantseva},
  {Hofverberg}, {Koldashov}, {Krutkov}, {Kvashnin}, {Leonov}, {Malvezzi},
  {Marcelli}, {Menn}, {Mikhailov}, {Mocchiutti}, {Orsi}, {Osteria}, {Papini},
  {Pearce}, {Picozza}, {Ricci}, {Ricciarini}, {Simon}, {Sparvoli},
  {Spillantini}, {Stozhkov}, {Vacchi}, {Vannuccini}, {Vasilyev}, {Voronov},
  {Yurkin}, {Zampa}, {Zampa}, and {Zverev}]{Adriani_Barbarino+09a}
{Adriani}, O.; {Barbarino}, G.C.; {Bazilevskaya}, G.A.; {Bellotti}, R.;
  {Boezio}, M.; {Bogomolov}, E.A.; {Bonechi}, L.; {Bongi}, M.; {Bonvicini}, V.;
  {Bottai}, S.; et al.
\newblock {An anomalous positron abundance in cosmic rays with energies
  1.5-100GeV}.
\newblock {\em \nat} {\bf 2009}, {\em 458},~607--609.

\bibitem[{Blasi} and {Amato}(2011)]{Blasi_Amato11a}
{Blasi}, P.; {Amato}, E.
\newblock {Positrons from pulsar winds}.
\newblock {\em  Astrophys. Space Sci. Proc.} {\bf 2011}, {\em
  21},~623--641.

\bibitem[{Weisskopf} \em{et~al.}(2000){Weisskopf}, {Hester}, {Tennant},
  {Elsner}, {Schulz}, {Marshall}, {Karovska}, {Nichols}, {Swartz},
  {Kolodziejczak}, and {O'Dell}]{Weisskopf_Hester+00a}
{Weisskopf}, M.C.; {Hester}, J.J.; {Tennant}, A.F.; {Elsner}, R.F.; {Schulz},
  N.S.; {Marshall}, H.L.; {Karovska}, M.; \mbox{{Nichols}, J.S.;} {Swartz}, D.A.;
  {Kolodziejczak}, J.J.; et al.
\newblock {Discovery of Spatial and Spectral Structure in the X-Ray Emission
  from the Crab Nebula}.
\newblock {\em Astrophys. J. Lett.} {\bf 2000}, {\em 536},~L81--L84.

\bibitem[{Pavlov} \em{et~al.}(2001){Pavlov}, {Kargaltsev}, {Sanwal}, and
  {Garmire}]{Pavlov_Kargaltsev+01a}
{Pavlov}, G.G.; {Kargaltsev}, O.Y.; {Sanwal}, D.; {Garmire}, G.P.
\newblock {Variability of the Vela Pulsar Wind Nebula Observed with Chandra}.
\newblock {\em Astrophys. J. Lett.} {\bf 2001}, {\em 554},~L189--L192.

\bibitem[{Gaensler} \em{et~al.}(2002){Gaensler}, {Arons}, {Kaspi},
  {Pivovaroff}, {Kawai}, and {Tamura}]{Gaensler_Arons+02a}
{Gaensler}, B.M.; {Arons}, J.; {Kaspi}, V.M.; {Pivovaroff}, M.J.; {Kawai}, N.;
  {Tamura}, K.
\newblock {Chandra Imaging of the X-Ray Nebula Powered by Pulsar B1509-58}.
\newblock {\em Astrophys. J.} {\bf 2002}, {\em 569},~878--893.

\bibitem[{Komissarov} and {Lyubarsky}(2004)]{Komissarov_Lyubarsky04a}
{Komissarov}, S.S.; {Lyubarsky}, Y.E.
\newblock {Synchrotron nebulae created by anisotropic magnetized pulsar winds}.
\newblock {\em Mon. Not. R. Astron. Soc.} {\bf 2004}, {\em 349},~779--792.

\bibitem[{Del Zanna} \em{et~al.}(2004){Del Zanna}, {Amato}, and
  {Bucciantini}]{Del-Zanna_Amato+04a}
{Del Zanna}, L.; {Amato}, E.; {Bucciantini}, N.
\newblock {Axially symmetric relativistic MHD simulations of Pulsar Wind
  Nebulae in Supernova Remnants. On the origin of torus and jet-like features}.
\newblock {\em Astron. Astrophys.} {\bf 2004}, {\em 421},~1063--1073.

\bibitem[{Olmi} \em{et~al.}(2016){Olmi}, {Del Zanna}, {Amato}, {Bucciantini},
  and {Mignone}]{Olmi_Del-Zanna+16a}
{Olmi}, B.; {Del Zanna}, L.; {Amato}, E.; {Bucciantini}, N.; {Mignone}, A.
\newblock {Multi-D magnetohydrodynamic modelling of pulsar wind nebulae: Recent
  progress and open questions}.
\newblock {\em J. Plasma Phys.} {\bf 2016}, {\em 82},~635820601.

\bibitem[{Olmi} \em{et~al.}(2015){Olmi}, {Del Zanna}, {Amato}, and
  {Bucciantini}]{Olmi_Del-Zanna+15a}
{Olmi}, B.; {Del Zanna}, L.; {Amato}, E.; {Bucciantini}, N.
\newblock {Constraints on particle acceleration sites in the Crab nebula from
  relativistic magnetohydrodynamic simulations}.
\newblock {\em Mon. Not.  R. Astron. Soc.} {\bf 2015}, {\em 449},~3149--3159.

\bibitem[{Woltjer}(1972)]{Woltjer72a}
{Woltjer}, L.
\newblock {Supernova Remnants}.
\newblock {\em Annu. Rev. Astron. Astrophys.} {\bf 1972}, {\em 10},~129--158.

\bibitem[{Reynolds}(2008)]{Reynolds08a}
{Reynolds}, S.P.
\newblock {Supernova Remnants at High Energy}.
\newblock {\em Annu. Rev. Astron. Astrophys.} {\bf 2008}, {\em 46},~89--126.

\bibitem[{Reynolds}(2011)]{Reynolds11a}
{Reynolds}, S.P.
\newblock {Particle acceleration in supernova-remnant shocks}.
\newblock {\em Astrophys. Space Sci.} {\bf 2011}, {\em 336},~257--262.

\bibitem[{Bell}(2004)]{Bell04a}
{Bell}, A.R.
\newblock {Turbulent amplification of magnetic field and diffusive shock
  acceleration of cosmic rays}.
\newblock {\mbox{\em Mon. Not.  R. Astron. Soc.}} {\bf 2004}, {\em 353},~550--558.

\bibitem[{Reynolds} \em{et~al.}(2012){Reynolds}, {Gaensler}, and
  {Bocchino}]{Reynolds_Gaensler+12a}
{Reynolds}, S.P.; {Gaensler}, B.M.; {Bocchino}, F.
\newblock {Magnetic Fields in Supernova Remnants and Pulsar-Wind Nebulae}.
\newblock {\em \ssr} {\bf 2012}, {\em 166},~231--261.

\bibitem[{Matthews} \em{et~al.}(2017){Matthews}, {Bell}, {Blundell}, and
  {Araudo}]{Mattheus_Bell+17a}
{Matthews}, J.H.; {Bell}, A.R.; {Blundell}, K.M.; {Araudo}, A.T.
\newblock {Amplification of perpendicular and parallel magnetic fields by
  cosmic ray currents}.
\newblock {\em Mon. Not.  R. Astron. Soc.} {\bf 2017}, {\em 469},~1849--1860.

\bibitem[{Xu} and {Lazarian}(2017)]{Xu_Lazarian17a}
{Xu}, S.; {Lazarian}, A.
\newblock {Magnetic Field Amplification in Supernova Remnants}.
\newblock {\em Astrophys. J.} {\bf 2017}, {\em 850}, 126.

\bibitem[{Conway}(1971)]{Conway71a}
{Conway}, R.G.
\newblock {Radio Polarization of the Crab Nebula}.
\newblock  In \emph{The Crab Nebula, Proceedings of the International Astronomical Union / Union Astronomique Internationale (Symposium No. 46),  Jodrell Bank, UK,  5--7 August 1970}; {Davies}, R.D., {Graham-Smith}, F., Eds.; Springer: Dordrecht, The Netherlands, 1971; p. 292.

\bibitem[{Ferguson}(1973)]{Ferguson73a}
{Ferguson}, D.C.
\newblock {A Comparison of the Optical and Radio Polarization of the Crab
  Nebula Pulsar.}
\newblock  \emph{Bull. Am. Astron. Soc.}  \textbf{1973}, \emph{5},  425.

\bibitem[{Velusamy}(1985)]{Velusamy85a}
{Velusamy}, T.
\newblock {Structure of the Crab Nebula---Intensity and polarization at 20 CM}.
\newblock {\em Mon. Not.  R. Astron. Soc.} {\bf 1985}, {\em 212},~359--365.

\bibitem[{Aumont} \em{et~al.}(2010){Aumont}, {Conversi}, {Thum}, {Wiesemeyer},
  {Falgarone}, {Mac{\'{\i}}as-P{\'e}rez}, {Piacentini}, {Pointecouteau},
  {Ponthieu}, {Puget}, {Rosset}, {Tauber}, and {Tristram}]{Aumont_Conversi+10a}
{Aumont}, J.; {Conversi}, L.; {Thum}, C.; {Wiesemeyer}, H.; {Falgarone}, E.;
  {Mac{\'{\i}}as-P{\'e}rez}, J.F.; {Piacentini}, F.; {Pointecouteau}, E.;
  {Ponthieu}, N.; {Puget}, J.L.; et al.
\newblock {Measurement of the Crab nebula polarization at 90 GHz as a
  calibrator for CMB experiments}.
\newblock {\em Astron. Astrophys.} {\bf 2010}, {\em 514},~A70.

\bibitem[{Dodson} \em{et~al.}(2003){Dodson}, {Lewis}, {McConnell}, and
  {Deshpande}]{Dodson_Lewis+03a}
{Dodson}, R.; {Lewis}, D.; {McConnell}, D.; {Deshpande}, A.A.
\newblock {The radio nebula surrounding the Vela pulsar}.
\newblock {\mbox{\em Mon. Not.  R. Astron. Soc.}} {\bf 2003}, {\em 343},~116--124.

\bibitem[{Kothes} \em{et~al.}(2006){Kothes}, {Reich}, and
  {Uyan{\i}ker}]{Kothes_Reich+06a}
{Kothes}, R.; {Reich}, W.; {Uyan{\i}ker}, B.
\newblock {The Boomerang PWN G106.6+2.9 and the Magnetic Field Structure in
  Pulsar Wind Nebulae}.
\newblock {\em Astrophys. J.} {\bf 2006}, {\em 638},~225--233.

\bibitem[{Ma} \em{et~al.}(2016){Ma}, {Ng}, {Bucciantini}, {Slane}, {Gaensler},
  and {Temim}]{Ma_Ng+16a}
{Ma}, Y.K.; {Ng}, C.Y.; {Bucciantini}, N.; {Slane}, P.O.; {Gaensler}, B.M.;
  {Temim}, T.
\newblock {Radio Polarization Observations of the Snail: A Crushed Pulsar Wind
  Nebula in G327.1-1.1 with a Highly Ordered Magnetic Field}.
\newblock {\em Astrophys. J.} {\bf 2016}, {\em 820},~100.

\bibitem[{Blondin} \em{et~al.}(2001){Blondin}, {Chevalier}, and
  {Frierson}]{Blondin_Chevalier+01a}
{Blondin}, J.M.; {Chevalier}, R.A.; {Frierson}, D.M.
\newblock {Pulsar Wind Nebulae in Evolved Supernova Remnants}.
\newblock {\em Astrophys. J.} {\bf 2001}, {\em 563},~806--815.

\bibitem[{Bucciantini} \em{et~al.}(2003){Bucciantini}, {Blondin}, {Del Zanna},
  and {Amato}]{Bucciantini_Blondin03a}
{Bucciantini}, N.; {Blondin}, J.M.; {Del Zanna}, L.; {Amato}, E.
\newblock {Spherically symmetric relativistic MHD simulations of pulsar wind
  nebulae in supernova remnants}.
\newblock {\em Astron. Astrophys.} {\bf 2003}, {\em 405},~617--626.

\bibitem[{Kothes} \em{et~al.}(2008){Kothes}, {Landecker}, {Reich}, {Safi-Harb},
  and {Arzoumanian}]{Kothes_Landecker+08a}
{Kothes}, R.; {Landecker}, T.L.; {Reich}, W.; {Safi-Harb}, S.; {Arzoumanian},
  Z.
\newblock {DA 495: An Aging Pulsar Wind Nebula}.
\newblock {\em Astrophys. J.} {\bf 2008}, {\em 687},~516--531.

\bibitem[{Moran} \em{et~al.}(2013){Moran}, {Shearer}, {Mignani},
  {S{\l}owikowska}, {De Luca}, {Gouiff{\`e}s}, and
  {Laurent}]{Moran_Shearer+13b}
{Moran}, P.; {Shearer}, A.; {Mignani}, R.P.; {S{\l}owikowska}, A.; {De Luca},
  A.; {Gouiff{\`e}s}, C.; {Laurent}, P.
\newblock {Optical polarimetry of the inner Crab nebula and pulsar}.
\newblock {\em Mon. Not.  R. Astron. Soc.} {\bf 2013}, {\em 433},~2564--2575.

\bibitem[{Hester}(2008)]{Hester08a}
{Hester}, J.J.
\newblock {The Crab Nebula: An Astrophysical Chimera}.
\newblock {\em Annu. Rev. Astron. Astrophys.} {\bf 2008}, {\em 46},~127--155.

\bibitem[{Marubini} \em{et~al.}(2015){Marubini}, {Sefako}, {Venter}, and {de
  Jager}]{Marubini_Sefako+15a}
{Marubini}, T.E.; {Sefako}, R.R.; {Venter}, C.; {de Jager}, O.C.
\newblock {A search for optical counterparts of the complex Vela X system}.
\newblock \emph{arXiv}  \textbf{2015}, arXiv:1501.00278.

\bibitem[{Moran} \em{et~al.}(2014){Moran}, {Mignani}, and
  {Shearer}]{Moran_Mignani+14a}
{Moran}, P.; {Mignani}, R.P.; {Shearer}, A.
\newblock {HST optical polarimetry of the Vela pulsar and nebula}.
\newblock {\em Mon. Not.  R. Astron. Soc.} {\bf 2014}, {\em 445},~835--844.

\bibitem[{Dubner} and {Giacani}(2015)]{Dubner_Giancani15a}
{Dubner}, G.; {Giacani}, E.
\newblock {Radio emission from supernova remnants}.
\newblock {\em Astron. Astrophys. Rev.} {\bf 2015}, {\em 23},~3.

\bibitem[{Jun}(1998)]{Jun98a}
{Jun}, B.I.
\newblock {Interaction of a Pulsar Wind with the Expanding Supernova Remnant}.
\newblock {\em Astrophys. J.} {\bf 1998}, {\em 499},~282--293.

\bibitem[{Bucciantini} \em{et~al.}(2004){Bucciantini}, {Amato}, {Bandiera},
  {Blondin}, and {Del Zanna}]{Bucciantini_Amato+04a}
{Bucciantini}, N.; {Amato}, E.; {Bandiera}, R.; {Blondin}, J.M.; {Del Zanna},
  L.
\newblock {Magnetic Rayleigh-Taylor instability for Pulsar Wind Nebulae in
  expanding Supernova Remnants}.
\newblock {\em Astron. Astrophys.} {\bf 2004}, {\em 423},~253--265.

\bibitem[{Inoue} \em{et~al.}(2013){Inoue}, {Shimoda}, {Ohira}, and
  {Yamazaki}]{Inoue_Shimoda+13a}
{Inoue}, T.; {Shimoda}, J.; {Ohira}, Y.; {Yamazaki}, R.
\newblock {The Origin of Radially Aligned Magnetic Fields in Young Supernova
  Remnants}.
\newblock {\em Astrophys. J. Lett.} {\bf 2013}, {\em 772},~L20.

\bibitem[{West} \em{et~al.}(2017){West}, {Jaffe}, {Ferrand}, {Safi-Harb}, and
  {Gaensler}]{West_Jaffe+17a}
{West}, J.L.; {Jaffe}, T.; {Ferrand}, G.; {Safi-Harb}, S.; {Gaensler}, B.M.
\newblock {When Disorder Looks Like Order: A New Model to Explain Radial
  Magnetic Fields in Young Supernova Remnants}.
\newblock {\em Astrophys. J. Lett.} {\bf 2017}, {\em 849},~L22.

\bibitem[{Gaensler}(1998)]{Gaensler98a}
{Gaensler}, B.M.
\newblock {The Nature of Bilateral Supernova Remnants}.
\newblock {\em Astrophys. J.} {\bf 1998}, {\em 493},~781--792.

\bibitem[{Reynoso} \em{et~al.}(2013){Reynoso}, {Hughes}, and
  {Moffett}]{Reynoso_Hughes+13a}
{Reynoso}, E.M.; {Hughes}, J.P.; {Moffett}, D.A.
\newblock {On the Radio Polarization Signature of Efficient and Inefficient
  Particle Acceleration in Supernova Remnant SN 1006}.
\newblock {\em Astron. J.} {\bf 2013}, {\em 145},~104.

\bibitem[{Harvey-Smith} \em{et~al.}(2010){Harvey-Smith}, {Gaensler}, {Kothes},
  {Townsend}, {Heald}, {Ng}, and {Green}]{Harvey-Smith_Gaensler+10a}
{Harvey-Smith}, L.; {Gaensler}, B.M.; {Kothes}, R.; {Townsend}, R.; {Heald},
  G.H.; {Ng}, C.Y.; {Green}, A.J.
\newblock {Faraday Rotation of the Supernova Remnant G296.5+10.0: Evidence for
  a Magnetized Progenitor Wind}.
\newblock {\em Astrophys. J.} {\bf 2010}, {\em 712},~1157--1165.

\bibitem[{Gotthelf} \em{et~al.}(2001){Gotthelf}, {Koralesky}, {Rudnick},
  {Jones}, {Hwang}, and {Petre}]{Gotthelf_Koralesky+01a}
{Gotthelf}, E.V.; {Koralesky}, B.; {Rudnick}, L.; {Jones}, T.W.; {Hwang}, U.;
  {Petre}, R.
\newblock {Chandra Detection of the Forward and Reverse Shocks in Cassiopeia
  A}.
\newblock {\em Astrophys. J. Lett.} {\bf 2001}, {\em 552},~L39--L43.

\bibitem[{Tang} and {Chevalier}(2012)]{Tang_Chevalier12a}
{Tang}, X.; {Chevalier}, R.A.
\newblock {Particle Transport in Young Pulsar Wind Nebulae}.
\newblock {\em Astrophys. J.} {\bf 2012}, {\em 752},~83.

\bibitem[{B{\"u}hler} and {Blandford}(2014)]{Buhler_Blandford14a}
{B{\"u}hler}, R.; {Blandford}, R.
\newblock {The surprising Crab pulsar and its nebula: A review}.
\newblock {\em Rep. Prog. Phys.} {\bf 2014}, {\em 77},~066901.

\bibitem[{Zrake} and {Arons}(2017)]{Zrake_Arons16a}
{Zrake}, J.; {Arons}, J.
\newblock {Turbulent Magnetic Relaxation in Pulsar Wind Nebulae}.
\newblock {\em Astrophys. J.} {\bf 2017}, {\em 847},~57.

\bibitem[{Tanaka} and {Asano}(2017)]{Tanaka_Asano16a}
{Tanaka}, S.J.; {Asano}, K.
\newblock {On the Radio-emitting Particles of the Crab Nebula: Stochastic
  Acceleration Model}.
\newblock {\em Astrophys. J.} {\bf 2017}, {\em 841},~78.

\bibitem[{Uzdensky} \em{et~al.}(2011){Uzdensky}, {Cerutti}, and
  {Begelman}]{Uzdensky_Cerutti+11a}
{Uzdensky}, D.A.; {Cerutti}, B.; {Begelman}, M.C.
\newblock {Reconnection-powered Linear Accelerator and Gamma-Ray Flares in the
  Crab Nebula}.
\newblock {\em Astrophys. J. Lett.} {\bf 2011}, {\em 737},~L40.

\bibitem[{Camus} \em{et~al.}(2009){Camus}, {Komissarov}, {Bucciantini}, and
  {Hughes}]{Camus_Komissarov+09a}
{Camus}, N.F.; {Komissarov}, S.S.; {Bucciantini}, N.; {Hughes}, P.A.
\newblock {Observations of `wisps' in magnetohydrodynamic simulations of the
  Crab Nebula}.
\newblock {\em Mon. Not.  R. Astron. Soc.} {\bf 2009}, {\em 400},~1241--1246.

\bibitem[{Sironi} and {Spitkovsky}(2011)]{Sironi_Spitkovsky11a}
{Sironi}, L.; {Spitkovsky}, A.
\newblock {Acceleration of Particles at the Termination Shock of a Relativistic
  Striped Wind}.
\newblock {\em Astrophys. J.} {\bf 2011}, {\em 741},~39.

\bibitem[{Mizuno} \em{et~al.}(2011){Mizuno}, {Lyubarsky}, {Nishikawa}, and
  {Hardee}]{Mizuno_Lyubarswky+11a}
{Mizuno}, Y.; {Lyubarsky}, Y.; {Nishikawa}, K.I.; {Hardee}, P.E.
\newblock {Three-dimensional Relativistic Magnetohydrodynamic Simulations of
  Current-driven Instability. II. Relaxation of Pulsar Wind Nebula}.
\newblock {\em Astrophys. J.} {\bf 2011}, {\em 728},~90.

\bibitem[{O'Neill} \em{et~al.}(2012){O'Neill}, {Beckwith}, and
  {Begelman}]{O'neill_Beckwith+12a}
{O'Neill}, S.M.; {Beckwith}, K.; {Begelman}, M.C.
\newblock {Local simulations of instabilities in relativistic jets---I.
  Morphology and energetics of the current-driven instability}.
\newblock {\em Mon. Not.  R. Astron. Soc.} {\bf 2012}, {\em 422},~1436--1452.

\bibitem[{Bandiera} and {Petruk}(2016)]{Bandiera_Petruk16a}
{Bandiera}, R.; {Petruk}, O.
\newblock {Radio polarization maps of shell-type supernova remnants---I.
  Effects of a random magnetic field component and thin-shell models}.
\newblock {\em Mon. Not.  R. Astron. Soc.} {\bf 2016}, {\em 459},~178--198.

\bibitem[{Petruk} \em{et~al.}(2017){Petruk}, {Bandiera}, {Beshley}, {Orlando},
  and {Miceli}]{Petruk_Bandiera+17a}
{Petruk}, O.; {Bandiera}, R.; {Beshley}, V.; {Orlando}, S.; {Miceli}, M.
\newblock {Radio polarization maps of shell-type SNRs---II. Sedov models with
  evolution of turbulent magnetic field}.
\newblock {\em Mon. Not.  R. Astron. Soc.} {\bf 2017}, {\em 470},~1156--1176.

\bibitem[{Bucciantini} \em{et~al.}(2017){Bucciantini}, {Bandiera}, {Olmi}, and
  {Del Zanna}]{Bucciantini_Bandiera+17a}
{Bucciantini}, N.; {Bandiera}, R.; {Olmi}, B.; {Del Zanna}, L.
\newblock {Modeling the effect of small-scale magnetic turbulence on the X-ray
  properties of Pulsar Wind Nebulae}.
\newblock {\em Mon. Not.  R. Astron. Soc.} {\bf 2017}, {\em 470},~4066--4074.

\bibitem[{Ng} and {Romani}(2004)]{Ng_Romani04a}
{Ng}, C.Y.; {Romani}, R.W.
\newblock {Fitting Pulsar Wind Tori}.
\newblock {\em Astrophys. J.} {\bf 2004}, {\em 601},~479--484.

\bibitem[{Bykov} \em{et~al.}(2009){Bykov}, {Uvarov}, {Bloemen}, {den Herder},
  and {Kaastra}]{Bykov_Uvarov+09a}
{Bykov}, A.M.; {Uvarov}, Y.A.; {Bloemen}, J.B.G.M.; {den Herder}, J.W.;
  {Kaastra}, J.S.
\newblock {A model of polarized X-ray emission from twinkling synchrotron
  supernova shells}.
\newblock {\em Mon. Not.  R. Astron. Soc.} {\bf 2009}, {\em 399},~1119--1125.

\bibitem[{Bykov} \em{et~al.}(2011{\natexlab{a}}){Bykov}, {Ellison}, {Osipov},
  {Pavlov}, and {Uvarov}]{Bykov_Ellison+11a}
{Bykov}, A.M.; {Ellison}, D.C.; {Osipov}, S.M.; {Pavlov}, G.G.; {Uvarov}, Y.A.
\newblock {X-ray Stripes in Tycho's Supernova Remnant: Synchrotron Footprints
  of a Nonlinear Cosmic-ray-driven Instability}.
\newblock {\em Astrophys. J. Lett.} {\bf 2011}, {\em 735},~L40.

\bibitem[{Bykov} \em{et~al.}(2011{\natexlab{b}}){Bykov}, {Osipov}, and
  {Ellison}]{Bykov_Osipov+11a}
{Bykov}, A.M.; {Osipov}, S.M.; {Ellison}, D.C.
\newblock {Cosmic ray current driven turbulence in shocks with efficient
  particle acceleration: The oblique, long-wavelength mode instability}.
\newblock {\em Mon. Not.  R. Astron. Soc.} {\bf 2011}, {\em 410},~39--52.

\bibitem[{Bucciantini}(2010)]{Bucciantini10a}
{Bucciantini}, N. {Polarization of pulsar wind nebulae}.
\newblock In {\em X-ray Polarimetry: A New Window in Astrophysics by Ronaldo
  Bellazzini, Enrico Costa, Giorgio Matt and Gianpiero Tagliaferri}; {Bellazzini}, R.,
  {Costa}, E., {Matt}, G., {Tagliaferri}, G., Eds.;~Cambridge
  University Press: Cambridge, UK, 2010;  p.~195.

\bibitem[{Soffitta} \em{et~al.}(2013){Soffitta}, {Barcons}, {Bellazzini},
  {Braga}, {Costa}, {Fraser}, {Gburek}, {Huovelin}, {Matt}, {Pearce},
  {Poutanen}, {Reglero}, {Santangelo}, {Sunyaev}, {Tagliaferri}, {Weisskopf},
  {Aloisio}, {Amato}, {Attin{\'a}}, {Axelsson}, {Baldini}, {Basso}, {Bianchi},
  {Blasi}, {Bregeon}, {Brez}, {Bucciantini}, {Burderi}, {Burwitz}, {Casella},
  {Churazov}, {Civitani}, {Covino}, {Curado da Silva}, {Cusumano}, {Dadina},
  {D'Amico}, {De Rosa}, {Di Cosimo}, {Di Persio}, {Di Salvo}, {Dovciak},
  {Elsner}, {Eyles}, {Fabian}, {Fabiani}, {Feng}, {Giarrusso}, {Goosmann},
  {Grandi}, {Grosso}, {Israel}, {Jackson}, {Kaaret}, {Karas}, {Kuss}, {Lai},
  {Rosa}, {Larsson}, {Larsson}, {Latronico}, {Maggio}, {Maia}, {Marin},
  {Massai}, {Mineo}, {Minuti}, {Moretti}, {Muleri}, {O'Dell}, {Pareschi},
  {Peres}, {Pesce}, {Petrucci}, {Pinchera}, {Porquet}, {Ramsey}, {Rea},
  {Reale}, {Rodrigo}, {R{\'o}{\.z}a{\'n}ska}, {Rubini}, {Rudawy}, {Ryde},
  {Salvati}, {de Santiago}, {Sazonov}, {Sgr{\'o}}, {Silver}, {Spandre},
  {Spiga}, {Stella}, {Tamagawa}, {Tamborra}, {Tavecchio}, {Teixeira Dias}, {van
  Adelsberg}, {Wu}, and {Zane}]{Soffitta_Barcons+13a}
{Soffitta}, P.; {Barcons}, X.; {Bellazzini}, R.; {Braga}, J.; {Costa}, E.;
  {Fraser}, G.W.; {Gburek}, S.; {Huovelin}, J.; {Matt}, G.; {Pearce}, M.;
  et al.
\newblock {XIPE: The X-ray imaging polarimetry explorer}.
\newblock {\em Exp. Astron.} {\bf 2013}, {\em 36},~523--567.

\bibitem[{Weisskopf} \em{et~al.}(2016){Weisskopf}, {Ramsey}, {O'Dell},
  {Tennant}, {Elsner}, {Soffitta}, {Bellazzini}, {Costa}, {Kolodziejczak},
  {Kaspi}, {Muleri}, {Marshall}, {Matt}, and {Romani}]{Weisskopf_Ramsey+16a}
{Weisskopf}, M.C.; {Ramsey}, B.; {O'Dell}, S.; {Tennant}, A.; {Elsner}, R.;
  {Soffitta}, P.; {Bellazzini}, R.; {Costa}, E.; {Kolodziejczak}, J.; {Kaspi},
  V.; et al.
\newblock {The Imaging X-ray Polarimetry Explorer (IXPE)}.
\newblock  In Proceedings of the Space Telescopes and Instrumentation 2016: Ultraviolet to Gamma Ray, Edinburgh, UK, 26 June--1 July 2016.

\bibitem[{Weisskopf} \em{et~al.}(1978){Weisskopf}, {Silver}, {Kestenbaum},
  {Long}, and {Novick}]{Weisskopf_Silver+78a}
{Weisskopf}, M.C.; {Silver}, E.H.; {Kestenbaum}, H.L.; {Long}, K.S.; {Novick},
  R.
\newblock {A precision measurement of the X-ray polarization of the Crab Nebula
  without pulsar contamination}.
\newblock {\em Astrophys. J. Lett.} {\bf 1978}, {\em 220},~L117--L121.

\bibitem[{Madsen} \em{et~al.}(2015){Madsen}, {Reynolds}, {Harrison}, {An},
  {Boggs}, {Christensen}, {Craig}, {Fryer}, {Grefenstette}, {Hailey},
  {Markwardt}, {Nynka}, {Stern}, {Zoglauer}, and {Zhang}]{Madsen_Reynolds+15a}
{Madsen}, K.K.; {Reynolds}, S.; {Harrison}, F.; {An}, H.; {Boggs}, S.;
  {Christensen}, F.E.; {Craig}, W.W.; {Fryer}, C.L.; {Grefenstette}, B.W.;
  {Hailey}, C.J.; et al.
\newblock {Broadband X-ray Imaging and Spectroscopy of the Crab Nebula and
  Pulsar with NuSTAR}.
\newblock {\em Astrophys. J.} {\bf 2015}, {\em 801},~66.

\bibitem[{Chauvin} \em{et~al.}(2016){Chauvin}, {Flor{\'e}n}, {Jackson},
  {Kamae}, {Kawano}, {Kiss}, {Kole}, {Mikhalev}, {Moretti}, {Olofsson},
  {Rydstr{\"o}m}, {Takahashi}, {Iyudin}, {Arimoto}, {Fukazawa}, {Kataoka},
  {Kawai}, {Mizuno}, {Ryde}, {Tajima}, {Takahashi}, and
  {Pearce}]{Chauvin_Floren+16a}
{Chauvin}, M.; {Flor{\'e}n}, H.G.; {Jackson}, M.; {Kamae}, T.; {Kawano}, T.;
  {Kiss}, M.; {Kole}, M.; {Mikhalev}, V.; {Moretti}, E.; {Olofsson}, G.;
  et al.
\newblock {Observation of polarized hard X-ray emission from the Crab by the
  PoGOLite Pathfinder}.
\newblock {\em Mon. Not.  R. Astron. Soc.} {\bf 2016}, {\em 456},~L84--L88.

\bibitem[{Chauvin} \em{et~al.}(2017){Chauvin}, {Flor{\'e}n}, {Friis},
  {Jackson}, {Kamae}, {Kataoka}, {Kawano}, {Kiss}, {Mikhalev}, {Mizuno},
  {Ohashi}, {Stana}, {Tajima}, {Takahashi}, {Uchida}, and
  {Pearce}]{Chauvin_Floren+17a}
{Chauvin}, M.; {Flor{\'e}n}, H.G.; {Friis}, M.; {Jackson}, M.; {Kamae}, T.;
  {Kataoka}, J.; {Kawano}, T.; {Kiss}, M.; \mbox{{Mikhalev}, V.; }{Mizuno}, T.;
 et al.
\newblock {Shedding new light on the Crab with polarized X-rays}.
\newblock {\em Sci. Rep.} {\bf 2017}, {\em 7},~7816.

\bibitem[{Forot} \em{et~al.}(2008){Forot}, {Laurent}, {Grenier},
  {Gouiff{\`e}s}, and {Lebrun}]{Forot_Laurent+08a}
{Forot}, M.; {Laurent}, P.; {Grenier}, I.A.; {Gouiff{\`e}s}, C.; {Lebrun}, F.
\newblock {Polarization of the Crab Pulsar and Nebula as Observed by the
  INTEGRAL/IBIS Telescope}.
\newblock {\em Astrophys. J. Lett.} {\bf 2008}, {\em 688},~L29.

\bibitem[{Moran} \em{et~al.}(2016){Moran}, {Kyne}, {Gouiff{\`e}s}, {Laurent},
  {Hallinan}, {Redfern}, and {Shearer}]{Moran_Kyne+16a}
{Moran}, P.; {Kyne}, G.; {Gouiff{\`e}s}, C.; {Laurent}, P.; {Hallinan}, G.;
  {Redfern}, R.M.; {Shearer}, A.
\newblock {A recent change in the optical and {$\gamma$}-ray polarization of
  the Crab nebula and pulsar}.
\newblock {\em Mon. Not.  R. Astron. Soc.} {\bf 2016}, {\em 456},~2974--2981.

\bibitem[{Vadawale} \em{et~al.}(2018){Vadawale}, {Chattopadhyay}, {Mithun},
  {Rao}, {Bhattacharya}, {Vibhute}, {Bhalerao}, {Dewangan}, {Misra}, {Paul},
  {Basu}, {Joshi}, {Sreekumar}, {Samuel}, {Priya}, {Vinod}, and
  {Seetha}]{Vadawale_Chattopadhyay+18a}
{Vadawale}, S.V.; {Chattopadhyay}, T.; {Mithun}, N.P.S.; {Rao}, A.R.;
  {Bhattacharya}, D.; {Vibhute}, A.; {Bhalerao}, V.B.; {Dewangan}, G.C.;
  {Misra}, R.; {Paul}, B.; et al.
\newblock {Phase-resolved X-ray polarimetry of the Crab pulsar with the
  AstroSat CZT Imager}.
\newblock {\em Nat. Astron.} {\bf 2018}, {\em 2},~50--55.

\bibitem[{de Ona Wilhelmi} \em{et~al.}(2017){de Ona Wilhelmi}, {Vink}, {Bykov},
  {Zanin}, {Bucciantini}, {Amato}, {Bandiera}, {Olmi}, {Uvarov}, and {XIPE
  Science Working Group}]{Wilhelmi_Vink+17a}
{de Ona Wilhelmi}, E.; {Vink}, J.; {Bykov}, A.; {Zanin}, R.; {Bucciantini}, N.;
  {Amato}, E.; {Bandiera}, R.; {Olmi}, B.; \mbox{{Uvarov}, Y.;} {XIPE Science Working
  Group}.
\newblock {Unveiling the magnetic structure of VHE SNRs/PWNe with XIPE, the
  x-ray imaging-polarimetry explorer}.
\newblock  In Proceedings of the 6th International Symposium on High Energy Gamma-Ray Astronomy, Heidelberg, Germany, 11--15 July 2016.

\end{thebibliography}
\end{document}